\documentstyle[epsf]{mn2e}

\title[Truncated disc versus extremely broad iron line]
{Truncated disc versus extremely broad iron line in XTE J1650--500}

\author[C. Done, M. ~Gierli\'nski]
{Chris Done$^1$ and Marek~Gierli\'nski$^{1,2}$\\
$^1$Department of Physics, University of Durham, South Road, Durham DH1 3LE,
UK\\
$^2$Obserwatorium Astronomiczne Uniwersytetu Jagiello{\'n}skiego, 30-244
Krak{\'o}w, Orla 171, Poland}

\date{Accepted for publication in MNRAS}
\pagerange{\pageref{firstpage}--\pageref{lastpage}}
\pubyear{2002}

\begin{document}

\topmargin = -0.5cm

\maketitle

\label{firstpage}

\begin{abstract}

There is growing evidence from both spectral and timing properties that
there is a truncated inner accretion disc in low mass accretion rate
Galactic black hole systems. The detection of {\em extremely} smeared
relativistic iron lines in some of these systems is the {\em only}
current piece of evidence which conflicts with this geometrical
interpretation of the low/hard state.  Here we show that the line width
in the {\it BeppoSAX\/} data of a bright low/hard state of the
transient black hole XTE~J1650--500 is indeed consistent with extreme
relativistic effects. However, the relativistic smearing can be {\em
significantly} reduced if there is also resonance iron K line {\em
absorption} from an outflowing disc wind. The iron line smearing is
then completely compatible with a truncated disc, so gives no
information on the black hole spin.

\end{abstract}


\section{Introduction}

The existence of a minimum stable orbit for material around a black
hole is a key prediction of general relativity in the strong field
limit.  For Schwarzschild black holes this is at $R_{\rm ms}=6R_g$
(where $R_g=GM/c^2$), while for maximally rotating Kerr black holes
(spin parameter $a=0.998$) this reduces to $1.23R_g$.  Bright accretion
flows give a way to observationally test such ideas as well as to
directly estimate the spin of the black hole. Optically thick material
in Keplerian orbits should emit a quasi-thermal spectrum with maximum
temperature emission from the largest luminosity/smallest area regions
i.e. set by the last stable orbit (Shakura \& Sunyaev 1973). As these
models are thermal, there is a clear prediction that their maximum
temperature, $T_{\rm max}$, should increase with total disc luminosity
as $L_{\rm disc}\propto T_{\rm max}^4$, if the emission is from a
constant area i.e. if there is a constant inner disc radius.  Strong
observational support for this comes from the Galactic black hole (GBH)
binary systems. These can show disc dominated spectra (often termed the
high/soft state) where $L_{\rm disc}\propto T_{\rm max}^4$ over large
changes in luminosity (Ebisawa et al. 1993; Kubota, Makishima \&
Ebisawa 2001; Kubota \& Makishima 2004). The fixed radius inferred from
the proportionality constant of this relation is consistent with a
Keplerian disc (stress-free inner boundary condition) in a
low--to--moderate spin spacetime ($a\la 0.8$) for all GBH for which
this experiment can be performed (Gierli{\'n}ski \& Done 2004; Li et
al. 2005; Shafee et al. 2005; Davis, Done \& Blaes 2005). This ties in
with the spins expected from models of stellar collapse. These predict
a maximum birth spin of $a\sim 0.8$ from the combination of slow
rotation of the pre-supernovae core and angular momentum losses through
gravitational wave radiation during collapse (e.g. Gammie, Shapiro \&
McKinney 2004), and there is not enough mass transfer in the binary
evolution to cause appreciable spin up from accretion (King \& Kolb
1996).

However, there is growing evidence that the accretion disc inner radius
can also be {\em larger} than $R_{\rm ms}$, due to the inner disc
evaporating into hot, geometrically thick accretion flow. These models
were developed in response to observations of GBH at low luminosity
(generally termed the low/hard state), where the expected accretion
disc emission is much weaker and at lower temperature than predicted by
an optically thick disc extending down to the last stable orbit (e.g
Esin et al. 2001; Frontera et al. 2001). Instead, the X-ray spectra
peak at $\sim$100~keV (see e.g. the review by Zdziarski \&
Gierli{\'n}ski 2004), requiring that the continuum is formed in hot,
optically thin plasma, which is not strongly cooled by Compton
scattering of seed photons (e.g. Zdziarski \& Gierli{\'n}ski 2004).
These cooling limits are very difficult (Barrio, Done \& Nayakshin
2003; though not impossible e.g. Beloborodov 1999) to reconcile with a
cool disc extending down to the minimum stable orbit. Other
observational advantages of truncated disc configuration is that it can
provide a geometric explanation for the major X-ray spectral state
transitions seen in GBH by a physical change in the accretion flow
structure. At low luminosities the inner cool disc is replaced by a hot
accretion flow. The transition radius between the two phases is far
from the hole, few photons from the disc illuminate the hot flow and
the spectrum is hard. As the transition radius moves inwards the disc
penetrates further into the hot flow, so more seed photons cool the
flow and its spectrum softens until the disc extends down to the last
stable orbit (Esin et al. 1997). This changing transition radius also
gives a changing size scale to associate with the changing
characteristic frequencies seen in the variability power spectra (see
e.g. the review by Done 2001).

Theoretically, such hot inner flows are generically produced by a range
of analytic approximations to the accretion flow equations. In these,
the protons are heated by gravity, while the electrons gain energy only
through Coulomb collisions and cool predominantly via Compton
scattering (Shapiro, Lightman \& Eardley 1975), advection (Narayan \&
Yi 1995), outflows (Blandford \& Begelman 1999), convection (Abramowicz
\& Igumenshchev 2001) or jets (Falcke, K{\"o}rding \& Markoff 2004).
Numerical simulations of the accretion flow at low densities, including
the self consistently generated magnetic dynamo viscosity (Balbus \&
Hawley 1991), appear to show some form of inner hot flow, though its
properties are more complex than any of the approximate analytic
solutions above (Hawley \& Balbus 2002).

Another argument for the truncated disc comes from the quasi-periodic
oscillations (QPOs) observed in the power density spectra of black hole
binaries. During the transition from the low/hard to the high/soft
spectral state their frequency typically increases (e.g. Cui et al.
1999; Rossi et al. 2004) indicating a decreasing size-scale in the
accretion flow. If the QPO frequency is directly related to the
transition radius between the cold disc and the hot inner flow (Stella
\& Vietri 1998; Titarchuk, Osherovich \& Kuznetsov 1999; Psaltis \&
Norman 2000) then its behaviour is consistent with a decreasing
truncation radius as the source approaches the soft state.

Clearly, the point at which the cool disc truncates is of great
importance to our understanding of the nature of the accretion flow,
and hence whether we can use observations of compact objects to test
strong gravity.  Fortunately there is an independent constraint on this
from the shape of the fluorescent iron line produced by X-ray
illumination of the cool disc. Special and general relativistic effects
are both important as the disc material is in very high velocity
(Keplerian) orbits in a strong gravitational field.  All these effects
get stronger as the disc extends closer to the black hole, so the width
of the line gives a measure of the inner edge of the cool fluorescing
material (Fabian et al. 1989; 2000). Thus for the accretion flow model
above in which there is a variable truncation radius between a disc and
inner hot flow, the line width should be correlated with spectral
hardness. Hard spectra (generally seen at low $L/L_{\rm Edd}$) should
have narrow lines, softer spectra should have broader lines. There are
indications that this correlation is seen both in AGN (Lubi{\'n}ski \&
Zdziarski 2001) and in GBH ({\.Z}ycki, Done \& Smith 1998; Churazov,
Gilfanov \& Revnivtsev 2000; Ibragimov et al. 2005).

However, some of the more recent results from the line shape seen in
the GBH do not fit easily into this picture. There are reported lines
{\em in the low/hard state} (i.e hard X-ray spectra, with QPO
frequencies which are not at their maximum values) which are so broad
as to require a disc around a maximally spinning black hole,
illuminated by highly centrally concentrated radiation field, perhaps
implying direct tapping of the spin energy of the black hole or at
least substantial continuous stress across the last stable orbit
(Miniutti, Fabian \& Miller 2004; Miller et al. 2002; Miller et al.
2004a). Plainly this would rule out the inner hot flow model if there
is no alternative explanation for the line width. The clearest example
of this discrepancy is the {\it BeppoSAX\/} data of the bright low/hard
state of XTE J1650--500. The line width in this spectrum is also in
conflict with the moderate spin derived from the luminosity-temperature
relation obtained when this object shows disc dominated spectra
(Gierli{\'n}ski \& Done 2004). Thus the extent of the disc inferred
from the iron line width in XTE J1650--500 seems at odds with the broad
band spectral shape, the QPOs, and the disc continuum emission.

Here we re-examine the {\it BeppoSAX\/} data on XTE J1650--500 and show
that while these data are indeed consistent with an extreme
relativistic line, they can be fit equally well (or better) using a
model of a `normal' relativistic line (i.e. from a truncated disc with
standard gravitational energy release), together with blueshifted
highly ionized narrow absorption features (predominantly resonance
lines of He- and H-like iron). These absorption features distort the
blue edge of the line, especially where the reflection spectrum is
ionized and/or from high inclination material. If the presence of such
material is confirmed then this removes the {\em only} observational
evidence against a truncated disc interpretation of the low/hard state
in GBH.

\section{The Data}

We extract the XTE~J1650--500 {\it BeppoSAX\/} archival data taken on
2001 September 11 for both the Medium-Energy Concentrator Spectrometer
(MECS) and Phoswich Detector System (PDS) instruments, using the
standard (September 1997) background, auxiliary response (MECS) and
redistribution files (MECS and PDS). We use the same energy ranges as
Miniutti et al. (2004), i.e. 2.5--10 keV (MECS) and 13--200 keV (PDS),
with 2 per cent systematic errors added in quadrature to the PDS data.
We checked the publicly available Crab PDS data and obtain reduced
$\chi^2/\nu = 0.9$ for a single power law of energy index 1.13, showing
that this prescription is an adequate description of the response
uncertainties.

\begin{figure*}
\begin{center}
\leavevmode \epsfxsize=0.8\textwidth \epsfbox{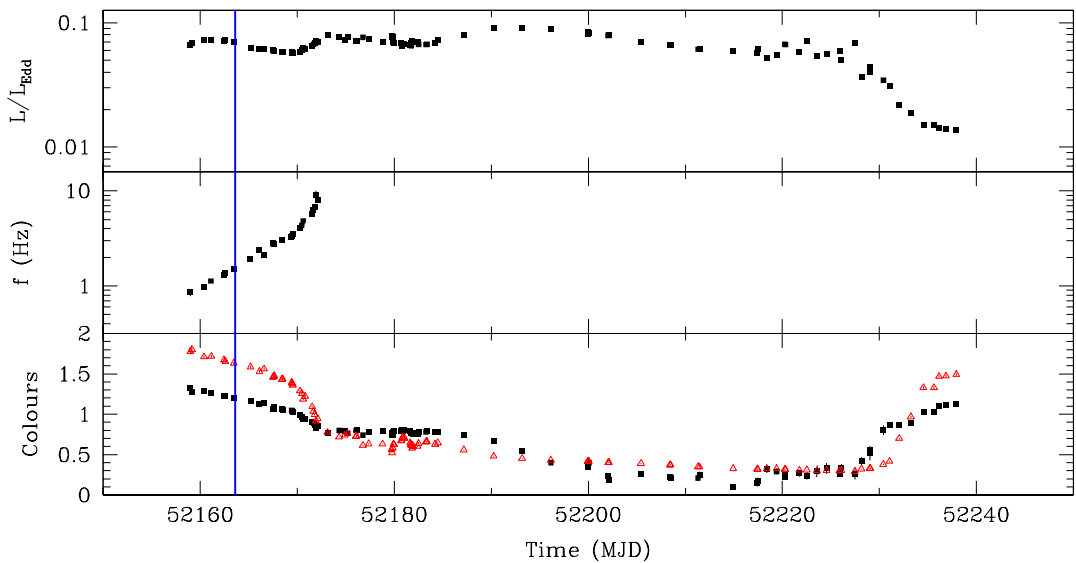}
\end{center}

\caption{The top panel shows the evolution of the total, absorption
corrected, bolometric luminosity (as a function of Eddington) for all
the {\it RXTE\/} pointed observations of the outburst of
XTE~J1650--500. The total flux is derived from fitting the PCA and
HEXTE data with a model of an accretion disc plus its Comptonized
emission (together with a broad line and smeared edge to mimic
reflection). The middle panel shows the low frequency QPO derived from
fitting the broad-band power spectra with multiple Lorentzian
components, while the bottom panel shows the soft (triangles, red in
colour) and hard (black squares) X-ray colours. Both QPO and colours
are still evolving smoothly at the point where the {\it BeppoSAX} data
were taken (vertical line). In the context of a truncated disc model
for the low/hard state, this implies that the disc has not yet reached
the last stable orbit.} \label{fig:xte}
\end{figure*}

We also extract the {\it RXTE} Proportional Counter Array (PCA) and
High-Energy X-ray Timing Experiment (HEXTE) spectra from the entire
outburst. We calculate the X-ray colours and estimate the bolometric
luminosity following the method of Done \& Gierli{\'n}ski (2003). We
used the distance of 4 kpc (Tomsick et al. 2003) and mass of 7.3
M$_\odot$ (Orosz et al. 2004) for the luminosity estimate. We also
extracted the power-density spectra from PCA observations in the 1/128
to 64 Hz frequency band and fitted them by a multiple Lorentzian model.

We use {\sc xspec} version 11.3.1 (Arnaud 1996) for spectral fitting.
The error of each model parameter is given for a 90 per cent confidence
interval. We fix the absorption column at $N_H=5\times 10^{21}$
cm$^{-2}$ in all fits (Miller et al. 2002; Miniutti et al. 2004).

\section{Timing and spectral evolution during the outburst}

Fig.~\ref{fig:xte} shows the estimated bolometric luminosity, the
frequency of the strong low-frequency QPO and X-ray colours obtained
from the {\it RXTE}. The vertical line marks the time of the {\it
BeppoSAX} observation analysed in this paper. The QPO frequency during
the {\it BeppoSAX} observation is plainly not the maximum seen from
this source. Any association of the QPO with the inner radius of the
cool disc clearly requires that the disc does {\em not} extend down to
the last stable orbit, but truncates {\em above} this point. The X-ray
colours show that the {\it BeppoSAX} data were taken while the source
is still smoothly softening, in a position which is associated with the
softest low/hard states. Given that the spectrum softens further {\em
after} the {\it BeppoSAX} observation, models of the broad band spectra
with a moving inner disc predict that the disc truncates {\em above}
the last stable orbit. Thus both the QPOs and broad band spectral shape
indicate that the innermost parts of a cool disc are not present. This
is in conflict with the extreme iron line smearing {\em in these same
data} which requires that the disc extends down to the last stable
orbit of a high spin black hole.

% all data and fits in /loc/done/done/sax1650
\section{MECS 2--10 keV data: $45^\circ$ inclination}

\subsection{Simple models}
\label{sec:simple}

%mecs_gio.xcm
With the MECS data we can reproduce the extreme relativistic broadening
of Miniutti et al. (2004). We use their phenomenological model of a
smeared edge and iron line with extreme Kerr spacetime distortions (the
{\sc laor} model in {\sc xspec}) on a continuum consisting of a disc
spectrum and power law. This gives an adequate fit ($\chi^2=70.9/65$)
to the data, and the parameters of the strong iron line (equivalent
width of $190_{-20}^{+30}$ eV) do indeed point to a very small inner
disc radius of $r_{\rm in} \equiv R_{\rm in}/R_g =
2.20_{-0.16}^{+0.06}$ with extreme central concentration of the
illumination, $\propto r^{-q}$ where $q=4.3^{+0.4}_{-0.3}$. A similarly
extreme line is seen using the same model in the {\it XMM-Newton} data
taken a few days after the {\it BeppoSAX\/} observation (Miller et al.
2002).

%mecs_pexriv.xcm
The {\sc xspec} {\sc laor} code is written for the specific case of
calculating the relativistic effects on a line which can be
approximated as an initial delta function. This is identical to the
transfer function, so relativistic effects on an arbitrary initial
spectrum can be calculated by simply convolving the initial spectrum
with this transfer function. We recoded the {\sc laor} model into a
convolution model (hereafter called {\sc conline}) in {\sc xspec} and
use this to smear a full reflected spectrum (the {\sc pexriv} model in
{\sc xspec} with additional narrow gaussian line) to replace the
phenomenological smeared edge/line features in the previous fit.
Convolution means that the spectrum outside of the observed energy band
can be important so we always extend the energy range used by {\sc
xspec} to 0.2--50~keV.

We follow Miller et al. (2002) and Miniutti et al. (2004) and fix the
inclination angle to $45^\circ$ and assume solar abundances.  We
constrain the radial emissivity index to be between 3 and 5 as
suggested by these previous fits. This gives $\chi^2=75.7/65$, i.e.
slightly worse than the phenomenological approximation for reflected
continuum emission. The best-fitting model is reflection dominated,
with $\Omega/2\pi=\infty$, i.e. for a model in which the intrinsic
continuum is hidden from view. However,  this is poorly constrained,
with the error range allowing $\Omega/2\pi > 0.7$. By contrast the line
parameters are fairly well constrained, with equivalent width of
$120\pm 25$~eV, and inner disc radius of $r_{\rm in} =
2.2^{+0.18}_{-0.12}$ and $q=3.7^{+0.3}_{-0.3}$. Thus the line
parameters do change with the addition of a reflected continuum, most
significantly in reducing the equivalent width, although the smearing
is also affected, with the illumination being not quite so centrally
concentrated. Detailed results on the line equivalent width (and to a
lesser extent the shape) derived without including a reflected
continuum should be treated with caution. Nonetheless, the small disc
inner radius is very similar to that in the phenomenological fits;
there is a very broad feature in the data between 4--7~keV as shown in
Fig.~\ref{fig:mecs}a.

\subsection{Complex reflection}
\label{sec:complex_reflection}

%mecs_ball2_hot_smeared_beta35.xcm: mecs_ball2_hot_beta35.xcm
The reflected component from the {\sc pexriv} fits in the previous
section is strongly ionized. This means that the intrinsic iron line
emission from the disc is {\em not} narrow. It should instead be
strongly Comptonized, and this Comptonization should also affect the
edge in the reflected continuum (Ross, Fabian \& Young 1999).  The
neglect of Compton scattering below 12~keV in {\sc pexriv} means that
this code has an edge which which is artificially deep and sharp for a
highly ionized disc. We replace the {\sc pexriv} reflected continuum
and narrow gaussian line with reflected spectra of D. Ballantyne from a
constant density ionized disc (hereafter CDID; Ballantyne, Iwasawa \&
Fabian 2001, based on Ross \& Fabian 1993) which are publicly available
as {\sc xspec} table models. These calculate the reflected spectrum
from an X-ray illuminated slab of constant density, including Compton
up- and down-scattering below 12~keV, with self consistent iron (and
other elements) line calculation for solar abundance material. These
give $\chi^2=93.7/67$ for a highly smeared, ionized slab, somewhat
worse than the (unphysical) fits above, but with two fewer free
parameters as the line energy and width are now calculated
self-consistently with the reflected continuum. The reflection emission
is highly ionized, $\lg (\xi / $erg cm s$^{-1}) =
3.79^{+0.03}_{-0.08}$, and again the spectrum is consistent with pure
reflected continuum, although the errors are wide and only restrict
$\Omega/2\pi>0.6$.

The best fit again picks out extreme relativistic smearing, with
$q=5^{+0}_{-1.2}$ and $r_{\rm in} = 3.0^{+0.3}_{-0.4}$. However, the
significance of detection of relativistic smearing is reduced with
respect to the reflection models which do not include Comptonization of
the line and edge features. The {\sc pexriv} based reflection models
give $\Delta\chi^2\sim 190$ for removing all smearing from the model,
while the {\sc CDID} reflection models give $\Delta\chi^2\sim 13$.

\begin{figure*}
\epsfxsize=0.98\textwidth \epsfbox{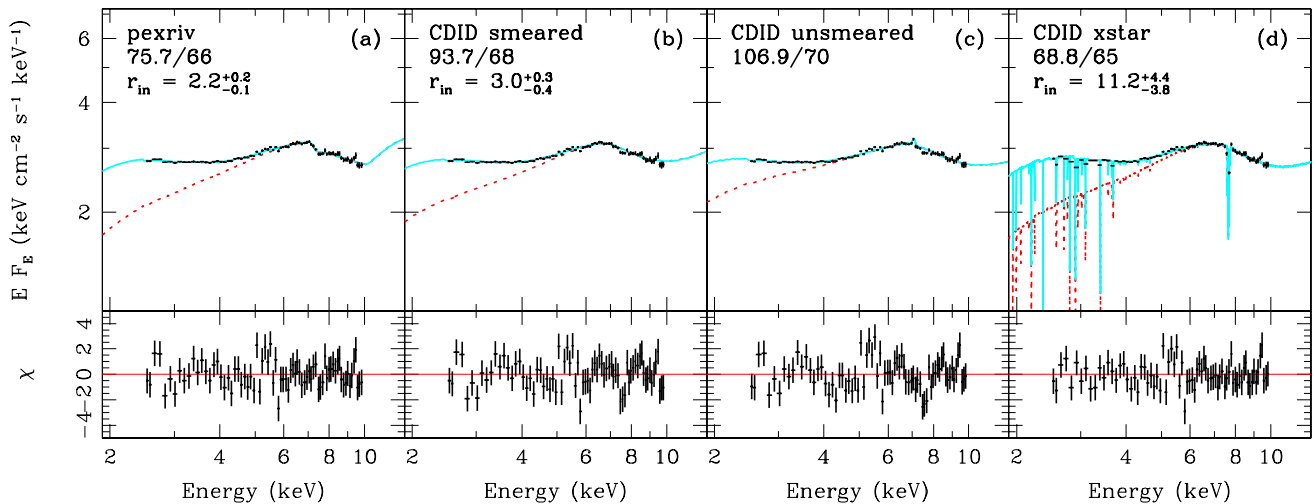}

\caption{MECS data fit to solar abundance reflection assuming an
inclination of $45^\circ$. The (red in colour) dotted curve corresponds
to the reflection, while the solid (cyan in colour) curve shows the sum
of the reflection and the disc emission which is out of scale in these
figures. Panel (a) shows the best fit using {\sc pexriv} based
reflection model, in which the reflector is highly ionized, and is
extremely relativistically smeared.  These relativistic effects
strongly distort the intrinsic reflected spectrum. Panel (b) shows the
very different reflected emission produced by {\sc CDID} models, where
the line and edge are intrinsically broad due to Comptonization. The
best-fitting reflected spectrum is highly ionized, and extremely
relativistically smeared. Panel (c) shows the same reflection model as
in (b) but with {\em no} relativistic effects at all applied, so this
fit is not dependent on inclination. The disc is again highly ionized,
so the line is intrinsically broad and this intrinsic shape is a fairly
good match to the data. This fit is only $\Delta\chi^2=13$ worse than
the best fit (with the loss of 2 degrees of freedom). Panel (d) shows
the fit including ionized {\em absorption} from outflowing material.
The inferred relativistic smearing reduces dramatically, so this fit is
relatively independent of inclination } \label{fig:mecs}
\end{figure*}

Figs.~\ref{fig:mecs}(b) and (c) show these fits with extreme
relativistic smearing, and no relativistic smearing, respectively.  It
is clear that the {\em intrinsic} reflected emission alone -- mainly
the edge in the reflected continuum rather than the line itself, as
there is very little line emission at these high ionizations -- can
match most of the broad feature in the data. This can have a strong
impact on the detection significance of relativistic smearing and some
impact on the detailed parameters.  Plainly it is important to use an
accurate description of ionized reflection. However, there is
considerable uncertainty in these models as they depend on the
(unknown) vertical and radial ionization structure of the disc. The
{\sc CDID} models assume that the reflecting material can be
approximated by a slab of constant (vertical and radial) density.  Yet
a disc is in hydrostatic equilibrium so its structure {\em responds} to
the incident radiation (Nayakshin, Kazanas \& Kallman 2000; Ballantyne,
Ross \& Fabian 2001). The upper layers are heated to the Compton
temperature of the illuminating flux, so they expand, becoming less
dense and more ionized. For hard power law illumination the Compton
temperature is large, so the upper layers are completely ionized,
forming a skin on top of the disc, but for softer illuminating spectra
(such as that of the data used here) the upper layers of the disc are
still visible so the difference between the hydrostatic and constant
ionization reflection models is much smaller (Done \& Nayakshin 2001).
As well as this vertical structure, there should also be a {\em radial}
dependance of the ionization structure of the disc, and the full
reflected emission should be the emissivity weighted integral of all of
these different radii regions with vertical ionization structure.

%mecs_xion_radial_05.xcm;
We use the {\sc xion} code (Nayakshin et al. 2000) to model reflection
from a {\em disc} of material in hydrostatic equilibrium from power law
illumination, i.e. including both radial and vertical ionization
gradients.  This can be used to produce the total emission (direct plus
reflected), in which the contribution of reflection is hardwired into
the spectrum for a given geometry. However, in order to give a
comparable number of degrees of freedom as in the {\sc CDID} fits, we
use {\sc xion} to calculate the reflected emission {\em only}, and add
a power law as the incident continuum so that the fraction
$\Omega/2\pi$ is allowed to vary. We use the magnetic flares geometry
to calculate the reflected emission from 10--100 Schwarzschild radii
from a disc accreting at $L/L_{\rm Edd}=0.05$ (see Fig.~\ref{fig:xte}).
The {\sc xion} code includes relativistic smearing, but only for
non-rotating black holes, so we switch this off and use the {\sc
conline} convolution model as before. This gives a somewhat worse fit
than before, with $\chi^2/\nu=100/67$. A better fit can be obtained by
setting $L/L_{\rm Edd}=1$, with $\chi^2/\nu=86.1/67$. This is due to
the higher disc density expected at higher mass accretion rates. Both
models give $r_{\rm in}\sim 2.2\pm 0.2$ and $q\sim 3.5\pm 0.3$ and
again the best fit is reflection dominated.

\subsection{Inclination}

The inclination of the system is is not well known, but recent studies
of the companion star indicate that its {\em minimum} value is $50\pm
3^\circ$ (Orosz et al. 2004), showing that the assumption of $45^\circ$
in the previous fits is probably too low.

\begin{figure}
\begin{center}
\leavevmode \epsfxsize=0.47\textwidth \epsfbox{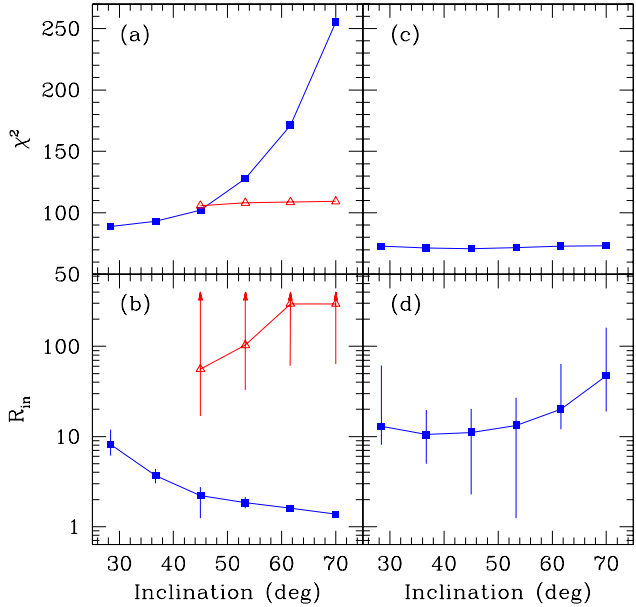}
\end{center}
\caption{Panel (a) shows the inclination dependance of $\chi^2$ for the
MECS data fit with a disc blackbody and relativistically smeared {\sc
CDID} ionized reflection with $q=3$. The best-fitting solution switches
between a highly smeared model (filled squares, blue in colour) and a
mostly unsmeared solution (open triangles, red in colour). Panel (b)
shows the derived disc inner radius for each solution. For the extreme
smearing, the increase in inclination causes a large blueshift of the
relativistic transfer function which is compensated by the fit by
increasing the gravitational redshift, i.e. decreasing the inner
radius. However, this also leads to increased broadening of the
spectral features, and the total spectral shape does not match the data
as well as at lower inclinations. Conversely, for the mostly unsmeared
fit, there is very little change with inclination in either $\chi^2$ or
derived radius. Panels (c) and (d) show how these results change with
the inclusion of outflowing ionized absorption (predominantly
blueshifted H- and He-like iron resonance lines). This gives better
fits at {\em all} inclinations, for a disc inner radius which is always
consistent with being larger than $6R_g$.} \label{fig:inc}
\end{figure}

The filled squares in Fig.~\ref{fig:inc}a show the increase in $\chi^2$
with inclination for the fits for the {\sc CDID} reflection models.
Plainly the lower inclination fits are preferred by the data. However,
we note that there is no minimum in $\chi^2$ at $45^\circ$ as seen from
{\sc pexriv} based reflection models (Miller et al. 2002), so removing
the only reason for using this inclination.  These fits become very
unstable so we fix $q=3$ and assume that the continuum is purely
reflected emission (i.e. no intrinsic continuum visible) in order to
produce the plots.  However, we checked that these assumptions do not
qualitatively affect the results, unless we allow for extremely
centrally concentrated emission. We can obtain good fits with $\chi^2
\sim 92$ at higher inclinations up to $70^\circ$, but only with
$q\sim15$.

At higher inclinations, the highly smeared solution becomes a worse fit
to the data because of the increasing blueshift of the relativistic
transfer function due to the larger projected line of sight velocities
(e.g. Fabian et al. 1989). The fit compensates for this by increasing
the gravitational redshift i.e. by decreasing the disc inner radius
and/or by increasing $q$. Since this figure is made for constant $q=3$
then this requires that the inner radius decreases. However, this does
not give enough gravitational redshift to compensate for the increased
line broadening and most of the increase in $\chi^2$ comes from this
increased blue extent of the spectral features.  These extremely
smeared fits then have a {\em worse} $\chi^2$ than the very different
fits in which the line is mostly {\em unsmeared}, as described in the
section above (see Fig.~\ref{fig:mecs}c). This solution is mostly
independent of inclination as the intrinsic shape of the reflected
emission below 10~keV is independent of inclination (Magdziarz \&
Zdziarski 1995), as are the relativistic effects for large inner disc
radii.  Because these two solutions, the extremely smeared and almost
unsmeared, are far apart in parameter space, they give a double minimum
in $\chi^2$ space. We show the $\chi^2$ for the mostly unsmeared fits
as the open triangles in Fig.~\ref{fig:inc}(a), and it is plain that
these are a much better description of the data for inclinations above
$45^\circ$ than the highly smeared fits (filled squares).
Fig.~\ref{fig:inc}(b) shows the corresponding disc inner radius
inferred for each solution as a function of inclination. Plainly, as
the best fit switches between these two very different solutions, there
is a corresponding dramatic jump in inferred disc inner radius.

The residuals to the (mostly) unsmeared fit for an inclination of
$45^\circ$ in Fig.~\ref{fig:mecs}(c) show a clear feature at
$\sim$~7.5~keV (the smeared fit in Fig.~\ref{fig:mecs}b also shows a
similar feature).  This looks like a fairly narrow absorption feature,
and including a narrow, gaussian line (with best-fitting energy
$7.58$~keV and equivalent width of $\sim$100~eV) improves the fit by
$\Delta\chi^2=17.4$, making it {\em significantly better}.  The
relativistic smearing reduces {\em dramatically}, now being entirely
consistent with $q=3$, and with a larger disc inner radius, $r_{\rm in}
= 9_{-4}^{+11}$.

\subsection{Narrow resonance line absorption}

Motivated by the fits above, we consider how such absorption lines can
occur.  Observationally, strong resonance absorption lines are often
seen from high inclination Galactic binary systems, generally from iron
K$\alpha$ He- and H-like iron at 6.7 and 7.0~keV (e.g. Ueda et al 1998;
Kotani et al. 2000; Lee et al. 2002; Ueda et al. 2004). However, if
this absorbing material is outflowing then substantial blueshifts could
be observed, and are seen in AGN (Pounds et al. 2003). If the feature
at $\sim 7.6$~keV is an absorption line from iron then this implies a
velocity of $\sim$0.09$c$ or 0.13$c$ depending on whether the
identification is with the H or He-like ion.

%mecs_ball2_xstar_i50.xcm, mecs_ball2_xstar_i70

We use the {\sc xstar} photoionization models which are publicly
available as {\sc xspec} multiplicative model tables. We choose the
ones calculated for power law illumination of a constant density
photoionized column with turbulent velocity of 100 km s$^{-1}$. This
gives an improved $\chi^2$ for {\em all} inclinations with the {\sc
CDID} reflection models (see Fig.~\ref{fig:inc}c), and dramatically
changes the inferred smearing of the reflected emission
(Fig.~\ref{fig:inc}d).  The relativistic line is now {\em always}
consistent with the illumination of $q=3$ expected from gravitational
energy release, and the inner disc radius is greater than $6R_g$.

The existence of several narrow features in the ionized absorber
between 6.6--7~keV means that $\chi^2$ space is complex, with multiple
local minima. However, the column in all fits is $\sim 10^{22}$
cm$^{-2}$ of ionized material ($\xi \sim 100$ erg cm s$^{-1}$)
outflowing at $\sim$0.1--0.15$c$, with the best-fitting redshift of
$0.151_{-0.003}^{+0.007}$. This model for an inclination of $45^\circ$
is shown in Fig.~\ref{fig:mecs}(d).

\begin{figure}
\begin{center}
\leavevmode \epsfxsize=0.35\textwidth \epsfbox{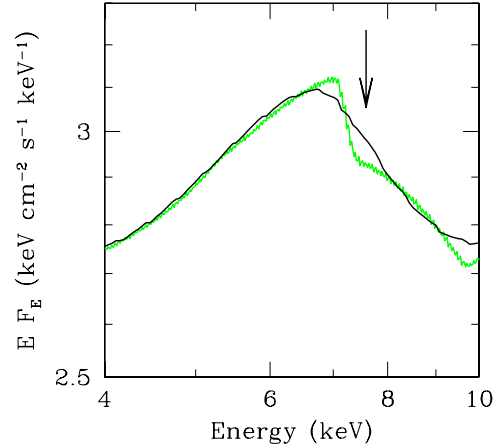}
\end{center}
\caption{The grey (green in colour) line shows the best-fitting extreme
smearing model where the intrinsic reflected emission is described by
{\sc pexriv} and a narrow gaussian line (Sec. \ref{sec:simple}). The
sharp edge to the transfer function gives the steep drop of the blue
wing of the iron line, while the deep edge in the ionised reflected
continuum is smoothed into a broad feature. There is a fairly steep
drop in the data at 7.2~keV so this model can match the curvature seen
(see Fig. \ref{fig:mecs}). The black line shows the best-fitting
extremely smeared {\sc CDID} model (Sec. \ref{sec:complex_reflection}).
The line and edge are now mostly blended together, intrinsically
broadened by Comptonisation in the intensely irradiated, hot upper
layers of the reflecting slab. Relativistic broadening further smooths
the spectrum, so there is no sharp drop at 7.5~keV. Since this drop is
present in the data, this model is a worse fit despite it being
physically more realistic. The arrow shows the energy inferred when
fitting an additional absorption line to the data. } \label{fig:mod}
\end{figure}

This effect is only seen when using the newer models of reflection from
ionized material, which include the self-consistent Comptonization of
the iron line and edge by the hot upper layers of the disc.
Fig.~\ref{fig:mod} (grey, or green in colour, line) shows the
best-fitting model using the {\sc pexriv} reflected continuum and
narrow Gaussian line, convolved with the {\sc laor} relativistic
profile. The sharp drop at $\sim$7.2~keV seen in the data can be
matched by the sharp blue wing of the relativistically smeared,
intrinsically {\em narrow} line. By contrast, the black line in
Fig.~\ref{fig:mod} shows the best-fitting model using the extremely
smeared, highly ionized {\sc CDID} reflection. The upper layer of the
irradiated reflecting material is hot, so Comptonizes much of the line
and edge, blending them together and making an intrinsically rather
smooth spectrum (Ross et al. 1999). Convolution of this spectrum with
relativistic effects makes it even broader, so there is no sharp change
in curvature produced by the blue wing of the line. Since the data does
have a sharp drop at 7.2--7.8~keV then these more physically self
consistent models of reflection give a worse fit. The significant
residuals in the spectrum at $\sim$7.5~keV then are better fit by
including outflowing absorption, predominantly resonance iron K lines
at rest frame energies of 6.7 and 6.95~keV. The drop seen in the data
then no longer drives the relativistic line profile to extreme
parameters and the derived smearing is completely compatible with a
truncated disc.

\subsection{Summary of MECS fits}

There is a clear broad residual associated with iron in these {\it
BeppoSAX\/} data from XTE J1650--500.  Simple models of reflection,
which ignore Comptonization of the line, require that the relativistic
effects are extreme, implying a spinning black hole and perhaps direct
extraction of the spin energy. However, these models also require that
the reflected emission is highly ionized. This means that
Comptonization of the line and edge is important, making these features
{\em intrinsically} broad and blended together. The strength and
significance of relativistic effects are then determined by the
detailed modelling of the intrinsic reflected emission, which in turn
depends on the (poorly known) vertical and radial structure of the
disc. Nevertheless, the current best models of ionized reflection
always pick out parameters where the innermost radius of the emitting
material is much closer than $6R_g$, and with radial emissivity rather
steeper than expected from gravitational energy release alone. However,
the smearing derived from these (best available) reflection models are
{\em not} robust to changes in inclination, nor to the addition of
ionized absorption features. While either (or both) of these effects
can resolve the discrepancy of the inner disc radius derived from the
line shape with that inferred from the overall evolution of spectral
and timing features, a further conflict remains which is that the fits
favour a reflection dominated solution.

\section{MECS and PDS 2--200~keV: Complex continuum models}

\subsection{Simple reflection models}

To get more constraints on the amount of reflection requires data at
higher energies, so we include the simultaneous PDS data to extend the
spectrum up to 200~keV. However, the spectrum is now much more
sensitive to details of the continuum modelling due to the wide
bandpass, and it is obvious that a power law can no longer be used.
Even an exponentially cutoff power law is not a good approximation to
the shape of a thermally Comptonized spectrum over such a wide
bandpass. The rollover at energies close to the electron temperature,
$kT_e$, is rather sharper than an exponential (Gierli{\'n}ski et al.
1997). Also, there are indications that the spectra can be more complex
than predicted by just a single temperature thermal Compton component,
with signatures of both thermal and non-thermal electrons even in the
low/hard state (McConnell et al. 2000; Ibragimov et al. 2005).

We use the {\sc eqpair} continuum model (Coppi 1999) in order to model
the curvature of the Comptonized continuum. This computes the
Comptonization of seed photons by a steady state electron distribution
which is {\em calculated} self-consistently from balancing an initial
heating rate with Compton and Coulomb cooling processes. The code
allows the input heating to be either non-thermal, with electrons
injected with a power-law distribution of $Q(\gamma)\propto
\gamma^{-s}$, between Lorentz factors $\gamma_{\rm min}$ and
$\gamma_{\rm max}$, or thermal, or some combination of both. However,
even if {\em all} the electron heating is non-thermal, the
self-consistent electron distribution (and hence output Comptonized
spectrum) can be mainly thermal with only a small non-thermal tail if
Coulomb collisions dominate the electron cooling. This happens where
the power injected in the hard electrons, $\ell_h$ (parameterized as a
compactness $\ell=L \sigma_T/R m_e c^3$; see e.g. Coppi 1999) is larger
than that injected in soft photons, $\ell_s$.  Conversely, where the
soft photons dominate then Compton cooling is most important, and the
electron distribution retains a mainly power-law shape with only a
small thermal component (Poutanen \& Coppi 1998).

As well as calculating the curvature of the Comptonized spectrum at
high energies, {\sc eqpair} also computes the low-energy cutoff due to
the seed photon distribution. This cutoff is typically at a few
$kT_{\rm seed}$, so is {\em within} the observed bandpass for the MECS
data. This curvature can make a large difference to the inferred disc
temperature and luminosity (see e.g. Done, {\.Z}ycki \& Smith 2002),
and could potentially distort the inferred spectrum in the red wing of
the iron line. We assume that the seed photons are the same as the
observed disc photons in all the fits below, and fix $\ell_s=10$.

%mecs_pds_eqpair_pexriv_i45u; mecs_pds_eqtherm_pexriv.xcm

This continuum complexity means that we cannot easily use the {\sc
CDID} reflection models since these are tabulated assuming a cutoff in
the continuum at a fixed energy of 100~keV. The assumed high energy
continuum shape affects the reflected emission at energies above $\sim
15$~keV due to the predominance of Compton down-scattering. Instead we
use the reflection model in {\sc eqpair} (which is based on {\sc
pexriv} but reflecting the self-consistent continuum from {\sc
eqpair}), together with a gaussian line. We relativistically smear both
reflected continuum and line with the {\sc laor} kernel, now extending
the bandpass for the convolution calculation from 0.2--1000~keV, and
assume an inclination of $45^\circ$.  Purely thermal electron heating
gives an (almost) acceptable fit, with $\chi^2_\nu=99.9/79$ for a
reflection dominated ($\Omega/2\pi=2.4^{+1.5}_{-0.8}$), highly smeared
fit ($q=3.8\pm0.2$, $r_{\rm in}=1.8^{+0.3}_{-0.1}$). However, purely
non-thermal injection models give a significantly better fit, with
$\chi^2=82.7/78$ (there is one extra free parameter which is the
spectral index of the power law electron injection). The subtle
difference in continuum curvature means that the solution is no longer
reflection dominated, with $\Omega/2\pi=0.72^{+0.12}_{-0.08}$ though it
is still strongly smeared ($q=3.5^{+0.3}_{-0.2}$ and $r_{\rm
in}=2.2^{+0.3}_{-0.2}$). The MECS and PDS spectrum and residuals for
this best-fitting non-thermal injection model are shown in
Fig.~\ref{fig:mecspds}(a).

\subsection{Complex reflection models}

Section \ref{sec:complex_reflection} showed that there are large
differences in the model spectra for ionized reflection.  While the
{\sc CDID} reflection models concentrate on the spectra below
$\sim$12~keV, computing the effects of photoionization and Compton up
and down-scattering at low energies, they do not accurately treat
Compton scattering at higher energies, nor include angle dependent
effects, nor allow for different illuminating continua.  This contrasts
with the {\sc pexriv} models of reflection which have very approximate
ionization balance (Done et al. 1992) and neglect Comptonization below
12~keV, but treat Comptonization at higher energies very accurately,
and give the reflected emission as a function of observed inclination
angle. {\sc pexriv} is based on the convolution kernels given by
Magdziarz \& Zdziarski (1995), so can be extended to work with {\em
any} illuminating continuum.

Plainly what is needed is a combination of these models, using the
better {\sc CDID} photoionization/Compton scattering of reflection at
low energies and the {\sc pexriv} accurate Compton down-scattering at
higher energies. The most flexible way to incorporate this is as a
convolution model in {\sc xspec}, so that the reflected spectrum can be
calculated for any arbitrary continuum. We have coded this by finding
the {\sc pexriv} cross-section at 10~keV which gives the same 12--14
keV reflected spectrum gradient as the {\sc CDID} models for a given
ionization parameter and illuminating spectral index. The corresponding
{\sc pexriv} Greens functions are used to form the spectrum above
12~keV. Below 12~keV the Greens functions are assumed to collapse to
delta functions, given by the ratio of reflected to incident flux in
the {\sc CDID} models, rescaled to match the normalization of {\sc
pexriv} at 14~keV.  This model is hereafter called {\sc refbal}.

\begin{figure*}
\begin{center}
\leavevmode \epsfxsize=0.73\textwidth \epsfbox{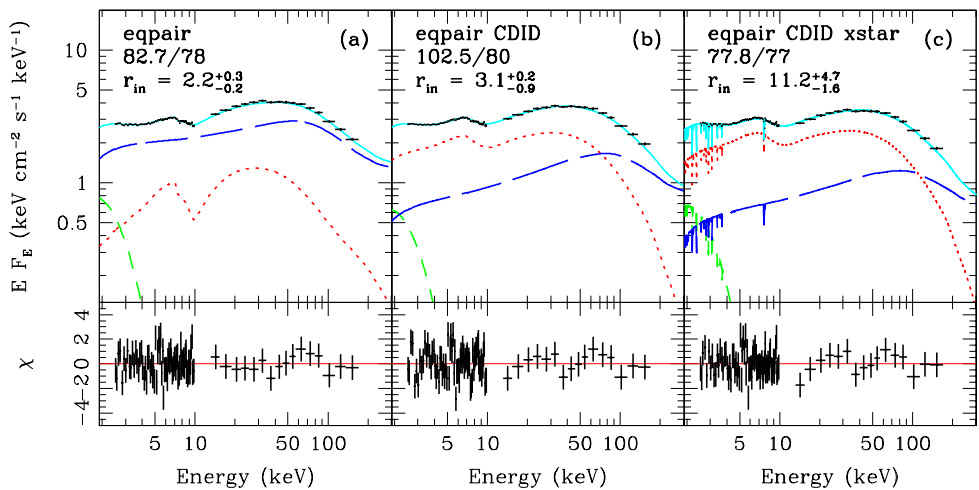}
\end{center}

\caption{MECS and PDS data fit with a disc blackbody (dashed curve,
green in colour) and {\sc eqpair} continuum (long-dashed, blue in
colour), together with the reflection (dotted, red in colour) from the
disc. Panel (a) shows the model with simple reflection, panel (b) with
CDID reflection and panel (c) the same, but with ionized blueshifted
absorption added.} \label{fig:mecspds}
\end{figure*}

%mecs_pds_eqpair_reflbal_i45u.xcm; mecs_pds_eqtherm_pexriv.xcm
We use this together with the {\sc eqpair} continuum model, and get a
somewhat worse fit than for the {\sc pexriv} models, with
$\chi^2=102.5/80$ for an illuminating continuum from non-thermal
injection. This is shown in Fig.\ref{fig:mecspds}(b) and by contrast
with the {\sc pexriv} reflection, it requires a larger fractional
contribution from reflection, with $\Omega/2\pi=2.0_{-0.7}^{+1.2}$.
However, both {\sc pexriv} and {\sc CDID} based models require that
reflection is highly smeared, with $r_{\rm in}$ = 2--3.

We then include the narrow absorption features from {\sc xstar} ionized
absorption.  As with the MECS data alone, this dramatically reduces the
derived relativistic smearing to being consistent with a disc truncated
{\em above} $r_{\rm in} =6$ for the {\sc CDID} reflection models.
However, the fit still has $\Omega/2\pi=2.4_{-0.4}^{+0.5}$ i.e. is
reflection dominated, although there is a significant contribution from
the intrinsic continuum component as well.

\subsection{Summary of MECS and PDS fits}

The inclusion of the PDS data up to 200~keV plainly gives more
constraints on the amount of reflection than the MECS data alone. With
data only up to 10~keV the amount of reflection is very poorly defined,
and the spectrum can be pure reflected emission with no intrinsic
continuum. Including the higher energy data shows that there is most
probably true continuum present, though the derived reflected fraction
is sensitive to details of how the reflected and continuum emission are
modelled. With the {\sc pexriv} based reflection models, the data are
consistent with $\Omega/2\pi <1$ as expected for a truncated disc,
while the more physical {\sc CDID} based reflection models have
substantially more reflection present, with $\Omega/2\pi\sim 2$. Both
these require extreme smearing and emissivity, but like the MECS-only
fits, the smearing is again sensitive to the inclusion of narrow
absorption lines from an ionized outflow for the {\sc CDID} reflection
models.

\section{Discussion}

The bright low/hard state spectrum seen by {\it BeppoSAX\/} from the
transient black hole XTE~J1650--500 is certainly consistent with
extremely smeared, reflection dominated emission as claimed by Miniutti
et al. (2004). However, it is also consistent with much less smearing
if there is some ionized, outflowing absorption which gives narrow
resonance lines at $\sim$7.5~keV. Reducing the amount of smearing makes
the reflected iron line consistent with the expectations of truncated
disc models, removing the conflict between the position of the inner
edge of the cool disc derived from the broad-band spectral shape and
QPOs. This is {\em directly testable} with high resolution data around
the iron line.

Such absorbing material is {\em often} seen in X-ray binaries,
especially those at fairly high inclination angles (Ueda et al. 1998;
Kotani et al. 2000; Lee et al. 2002; Ueda et al. 2004), including XTE
J1650--500 (Miller et al. 2004a). The one difference here is that the
outflowing material has velocity $\sim$4.5$\times10^4$~km s$^{-1}$,
much higher than the $\sim$500~km s$^{-1}$ typically seen in these
objects. A low velocity wind would be launched from fairly large radii
in the disc, as expected for line driving from the UV emitting region
of the GBH disc (Proga \& Kallman 2002). By contrast, the much higher
velocities required here would need to be associated with a launch from
the inner accretion disc, where line driving is not effective due to
the high ionization. This suggests rather that the wind would have to
be magnetically driven, and/or be associated with the jet. We note that
these data are taken shortly after the time of the maximum in radio
emission (Corbel et al. 2004), so the inferred very fast outflow could
be associated with the jet ejection.

We can estimate the mass outflow rate. For a constant-velocity outflow
the mass outflow rate at a radius $R$ is
\begin{equation}\dot{M}_{\rm out} = 4\pi R^2 n m_p f
v,\end{equation}where $n$ is the number density of the outflowing
material, $v$ is its velocity, $f \equiv \Omega/4\pi$ is the fractional
solid angle in which the outflow is directed and $m_p$ is the proton
mass. In a steady state $\dot{M}_{\rm out}$ is constant for all radii.
The column density measured by the observer is \begin{equation}N_H =
\int_{R_{\rm in}}^\infty n dR = {1 \over 4 \pi f m_p} {\dot{M}_{\rm
out} \over R_{\rm in} v},\end{equation} therefore
\begin{equation}\dot{M}_{\rm out} = 4\pi f m_p N_H R_{\rm in}
v.\end{equation} For $N_H = 10^{22}$ cm$^{-2}$, $R_{\rm in} = 10R_g$,
$v = 0.15c$ and $f \le 1$ we find $\dot{M}_{\rm out} \la
1.4\times10^{16}$ g s$^{-1} \approx 10^{-3}\dot{M}_{\rm Edd}$ (assuming
accretion efficiency of 0.1 and black hole mass of 10 M$_\odot$). This
is only about 1 per cent or less of estimated inflow mass accretion
rate and constitutes a very feasible number.

Our spectral fitting also shows that this absorption does {\em not}
make the data consistent with all the expectations of a truncated disc.
The best current models of reflection based on irradiation of a slab of
constant density give reflection fractions which are much larger than
expected for a truncated disc geometry. This is not necessarily the
case using the older {\sc pexriv} reflection models, which can give
much smaller reflection fractions. The difference is due to
Comptonization of the line and edge in the reflected emission as it
travels through the upper layers of the disc (Ross et al. 1999). This
effect is neglected in {\sc pexriv}, so the iron edge is rather sharp.
On the other hand, the {\sc CDID} reflection is not calculated for the
disc temperatures appropriate for Galactic binaries i.e. 0.5--1~keV. In
such hot material, collisional processes alone will ionize iron up to
the K shell (Davis et al. 2005). By contrast, the {\sc CDID} models do
not include collisional ionization, and thus require intense
illumination to ionize iron up to the K shell to match the data. This
pulls the temperature of the upper layers to the Compton temperature of
1.5--2~keV, for spectra with $\Gamma\sim$ 1.9--2 (Ross et al. 1999;
Nayakshin et al. 2000). This is a factor 2--3 larger than the observed
disc temperature in these data. Thus while the {\sc pexriv} models
underestimate the width of the line and edge by having no Compton
scattering below 12~keV, the {\sc CDID} models may overestimate it for
Galactic binaries. From our spectral fits, the derived reflected
fraction is sensitive to the detailed shape of this edge. Better models
of reflection for conditions appropriate to Galactic binary discs will
enable this effect to be quantified, removing this as a source of
uncertainty.

The amount of reflection is also sensitive to the detailed form assumed
for the continuum. This is not easily defined even with broad bandpass
data. Obviously, the wider the bandpass, the more constraints there are
on the form of the continuum curvature, but also the more the spectrum
is sensitive to the detailed form of continuum assumed. Broad bandpass
data suggest that the spectrum has convex curvature which can be
modelled by an additional soft excess component (e.g. Di Salvo et al.
2001; Ibragimov et al. 2005). This makes the intrinsic continuum at
higher energies rather harder than expected from the observed 2--10~keV
emission, reducing the amount of reflection required to match the data
at high energies. Including such a component in the model makes the
reflected fraction almost completely undetermined and consistent with a
truncated disc with the {\sc refbal} and ionized absorber models. Such
additional continuum complexity is not required statistically from the
spectral data alone (even with the moderate spectral resolution and
broad bandpass of {\it BeppoSAX}). However, energy resolved {\em
variability} could give additional constraints as such data can fairly
unambiguously pinpoint the seed photon energy seen by the Comptonizing
plasma (Gierli{\'n}ski \& Zdziarski 2005), so could disentangle the
shape of the soft emission.

\section{Other extreme line objects}

Most detections of extreme broad iron lines in GBH are in the very high
state (alternatively named the steep power law state) rather than in
the low/hard state examined here (see e.g. Miller et al. 2004a). This
(typically rather high mass accretion rate) spectral state shows strong
disc emission as well as a strong, steep high-energy tail, while the
power spectrum of the variability shows a strong QPO which is close to
its maximum frequency (e.g. McClintock \& Remillard 2006). Thus both
spectral and timing properties imply that the inner disc does extend
close to the black hole (quite how close is difficult to determine due
to the complexity of the spectra: Kubota \& Done 2004; Done \& Kubota
2005). Thus there are no obvious conflicts as for the low/hard state.
There are objects where the disc dominated spectra imply moderate spin
(from the $L_{\rm disc}\propto T_{\rm max}^4$ relation) while the iron
line smearing implies extreme spin but these could be explained by the
accretion flow extending below $R_{\rm ms}$ in the very high state
(Reynolds \& Begelman 1997).

However, there {\em is} a more subtle conflict with the extreme line in
these data which is that the very high state spectra typically show
strongly Comptonized emission.  The broad band spectral shape seen in
e.g. the simultaneous {\it ASCA}-{\it RXTE}-OSSE data from
XTE~J1550-564 implies that the inner disc is almost completely covered
by material with optical depth 2--3 (Done \& Kubota 2005;
Gierli{\'n}ski \& Done 2003). Thus the spectral shape implies that very
little of the inner disc is seen directly, so it {\em cannot} make much
contribution to the line emission. Yet the {\em same} {\it ASCA} data
require an extreme broad line when fit by {\sc pexriv} based models
(Miller et al. 2004b). Again, we suggest that resonance line absorption
in an outflowing wind can modify the derived line width and remove
these conflicts in the very high state spectra. To test this, we fit
the publicly available standard products Gas Imaging Spectrometer data,
as this does not suffer from the pileup problems of the Solid-State
Imaging Spectrometer for such a bright source. The relativistic
smearing required for a {\sc pexriv} reflected continuum and narrow
line are extreme, with $r_{in}=1.6_{-0.4}^{+0.3}$ and $q=3.7\pm0.3$,
with $\chi^2$ = 796.5/742, similar to the parameters found by Miller et
al. (2004b). Replacing this with {\sc CDID} reflection gives similar
extreme relativistic smearing but a worse fit ($\chi^2$ = 824.6/744).
Including {\sc xstar} ionized absorption improves the fit ($\chi^2$ =
805.4/741), and as in XTE~J1650--500 this reduces the relativistic
effects to the point where $r_{in}=31_{-15}$ (no upper limit) with $q$
fixed at 3.

The {\em extreme} broad lines inferred in some AGN are similarly
potentially affected by absorption complexity changing the line
parameters. Here there are as yet no obvious problems with this
interpretation, especially as extreme spin might be {\em expected}
given the hierarchical growth of black holes through accretion
(Volonteri et al. 2005).  Nonetheless, if the apparently extreme GBH
iron lines are actually consistent with low spin, normal illumination
rather than extraction of rotational energy from an extreme spin black
hole, then it is likely that the AGN discs behave similarly. We note
that the recent {\it Chandra} spectrum of the most secure extreme
relativistic line, that in MCG--6-30-15, reveals the presence of
ionized absorption at a similar column and ionization as that required
here to reduce the relativistic smearing to levels consistent with a
truncated disc (Young et al. 2005). Young et al. (2005) checked that
this absorption did not change the extreme smearing deduced from a {\sc
laor} description of the line, but this is unsurprising due to the
unphysically sharp blue wing of the line produced in such models (see
Fig.~\ref{fig:mod}). Our fitting showed that the relativistic line
parameters only changed significantly when using the {\sc CDID} highly
ionized reflection models as these have a much smoother intrinsic
spectrum around the iron line/edge due to Comptonization. We urge
testing the robustness of the extreme relativistic effects to the
inclusion of ionized absorption with more physical ionized reflection
models.

\section{Conclusions}

The truncated disc/hot inner flow geometry is very successful in
qualitatively (and sometimes quantitatively) explaining many disparate
properties of the low/hard state in GHB. If the radius at which the hot
flow evaporates from cool disc progressively decreases as the source
approaches the transition to the soft state then this can account for
the observed evolution in spectral and timing properties. However, if
there really are detections of substantial amounts of iron line
emission from cool disc-like material down at the last stable orbit
{\em in the low/hard state}, then the truncated disc idea is simply
wrong. Some of the extreme line emission might be explained by clumps
from a disrupted disc embedded in the hot flow, but this would give
rise to only a small amount of line since the covering fraction of the
clumps would be rather less than from a disc, and these are embedded in
a hot flow with optical depth of around unity, so Compton scattering
suppresses the line.

We show that the bright low/hard state spectrum seen by {\it BeppoSAX}
from the transient black hole XTE~J1650--500 can indeed be interpreted
as requiring strong, extremely smeared reflected emission from a
disc-like structure extending down to $\sim 2R_g$ as shown by Miniutti
et al. (2004). These data clearly illustrate the conflict between this
and the position of the inner edge of the cool disc derived from the
broad band spectral shape and QPOs.  The small inner radius inferred
from the line width also conflicts (though more weakly) with the
moderate spin inferred both from theoretical models of stellar collapse
and from the observed $L_{\rm disc}\propto T_{\rm max}^4$ relation in
this object.  We show that {\em all} these conflicts can be removed if
relativistic effects on the line and reflected continuum are
overestimated due to the presence of absorption line features in the
spectrum.  The derived line width is then easily consistent with normal
emissivity and a disc truncated at $\sim$10--20 $R_g$ (so giving no
information on the black hole spin). The reflection feature are most
probably shaped by strong, but not necessarily extreme, gravity. We
suggest that this effect is robust, i.e. that moderate columns of
highly ionized absorbing material (predominantly resonance lines of H-
and He-like iron) can {\em always} reduce the inferred relativistic
smearing of a highly ionized reflected spectrum. This could then remove
the more subtle conflict seen in the very high state GBH spectra, where
the disc should be down close to the last stable orbit (so strong
relativistic smearing is expected) but the spectral shape shows that
the inner disc is mostly covered  by an optically thick, hot corona. We
stress that the presence of absorption at this level is directly {\em
testable} with high resolution data.

While there are no such direct conflicts for AGN, we note that the most
secure extreme line candidate, MCG--6-30-15, has highly ionized
reflection (e.g. Wilms et al. 2002) and a recent {\it Chandra} HETG
spectrum shows absorption at the same column and ionization (Young et
al. 2005) as inferred here for XTE~J1650--500. We urge a re-analysis of
these data to see whether the extreme line parameters are robust to the
presence of absorption when using the more physical reflection models
which include self-consistent Comptonization of the line and edge as
opposed to models in which the line is intrinsically narrow.

\section*{Acknowledgements}

MG and CD thank PPARC for support through a fellowship and senior
fellowship, respectively.

%-----------------------------------------------------

\label{lastpage}


\begin{thebibliography}{}



\bibitem[Abramowicz \& Igumenshchev(2001)]{2001ApJ...554L..53A} Abramowicz
M.~A., Igumenshchev I.~V.\ 2001, ApJ, 554, L53

\bibitem[]{arn96} Arnaud K. A., 1996, in Jacoby G. H., Barnes J., eds.,
Astronomical Data Analysis Software and Systems V. ASP Conf. Series
Vol.\ 101, San Francisco, p.\ 17

\bibitem[]{} Ballantyne D. R., Iwasawa K., Fabian, 2001, MNRAS, 323,
  506

\bibitem[]{} Ballantyne D. R., Ross R. R., Fabian A. C., 2001, MNRAS, 327, 10

\bibitem[]{} Barrio F.E., Done C., Nayakshin S., 2003, MNRAS, 342, 557

\bibitem[]{} Balbus S.A., Hawley J., 1991, ApJ, 376, 214

\bibitem[Beloborodov(1999)]{1999ApJ...510L.123B} Beloborodov A.~M.\ 1999,
ApJ, 510, L123

\bibitem[Blandford \& Begelman(1999)]{1999MNRAS.303L...1B} Blandford
R.~D., Begelman M.~C.\ 1999, MNRAS, 303, L1

\bibitem[]{} Churazov E., Gilfanov M., Revnivtsev M., 2000, preprint,
astro-ph/0002415

\bibitem[]{} Coppi P. S., 1999, in ASP Conf. Ser. 161, 375

\bibitem[Corbel et al.(2004)]{2004ApJ...617.1272C} Corbel S., Fender
R.~P., Tomsick J.~A., Tzioumis A.~K., Tingay S.\ 2004, ApJ, 617, 1272

\bibitem[Cui et al.(1999)]{1999ApJ...512L..43C} Cui W., Zhang S.~N.,
Chen W., Morgan E.~H.\ 1999, ApJ, 512, L43

\bibitem[]{} Davis S. W., Done C., Blaes O., 2005, in preparation

\bibitem[Di Salvo et al.(2001)]{2001ApJ...547.1024D} Di Salvo T., Done
C., {\.Z}ycki P.~T., Burderi L., Robba N.~R.\ 2001, ApJ, 547, 1024

\bibitem[\protect\citeauthoryear{Done}{2001}]{2001AdSpR..28..255D} Done C.,
2001, AdSpR, 28, 255

\bibitem[]{} Done C., Nayakshin S., 2001, ApJ, 546, 419

\bibitem[Done \& Gierli{\'n}ski(2003)]{2003MNRAS.342.1041D} Done C.,
Gierli{\'n}ski M.\ 2003, MNRAS, 342, 1041

\bibitem[]{} Done C., Kubota A.,  2005, in preparation

\bibitem[]{} Done C., {\.Z}ycki P., Smith D. A., 2002, MNRAS, 331, 453 %cygx2

\bibitem[]{} Done C., et al., 1992, ApJ, 395, 275

\bibitem[Ebisawa et al.(1993)]{1993ApJ...403..684E} Ebisawa K., Makino
F., Mitsuda K., Belloni T., Cowley A.~P., Schmidtke P.~C., Treves A.\
1993, ApJ, 403, 684

\bibitem[]{} Esin A. A., McClintock, J. E., Narayan, R. 1997, ApJ, 489, 865

\bibitem[Esin et al.(2001)]{2001ApJ...555..483E} Esin A.~A., McClintock
J.~E., Drake J.~J., Garcia M.~R., Haswell C.~A., Hynes R.~I., Muno
M.~P.\ 2001, ApJ, 555, 483

\bibitem[Fabian et al.(1989)]{1989MNRAS.238..729F} Fabian A.~C., Rees
M.~J., Stella L., White N.~E.\ 1989, MNRAS, 238, 729

\bibitem[]{} Fabian A. C. et al., 2000, PASP, 112, 1145

\bibitem[Falcke et al.(2004)]{2004A&A...414..895F} Falcke H., K{\"o}rding
E., Markoff S.\ 2004, A\&~A, 414, 895

\bibitem[Frontera et al.(2001)]{2001ApJ...561.1006F} Frontera F., et al.\
2001, ApJ, 561, 1006

\bibitem[Gammie et al.(2004)]{2004ApJ...602..312G} Gammie C.~F., Shapiro
S.~L., McKinney J.~C.\ 2004, ApJ, 602, 312

\bibitem[Gierli{\'n}ski \& Done(2004)]{2004MNRAS.347..885G} Gierli{\'n}ski
M., Done C.\ 2004, MNRAS, 347, 885

\bibitem[]{} Gierli{\'n}ski M., Zdziarski A.A., 2005, MNRAS, in press

\bibitem[\protect\citeauthoryear{Gierlinski et
al.}{1997}]{1997MNRAS.288..958G} Gierlinski M., Zdziarski A.~A., Done
C., Johnson W.~N., Ebisawa K., Ueda Y., Haardt F., Phlips B.~F., 1997,
MNRAS, 288, 958

\bibitem{} Hawley J. F., Balbus S. A., 2002, ApJ, 573, 736

\bibitem[Ibragimov et al.(2005)]{2005MNRAS.362.1435I} Ibragimov A.,
Poutanen J., Gilfanov M., Zdziarski A.~A., Shrader C.~R.\ 2005, MNRAS,
362, 1435

\bibitem[King \& Kolb(1999)]{1999MNRAS.305..654K} King A.~R., Kolb U.\
1999, MNRAS, 305, 654

\bibitem[Kotani et al.(2000)]{2000ApJ...539..413K} Kotani T., Ebisawa K.,
Dotani T., Inoue H., Nagase F., Tanaka Y., Ueda Y.\ 2000, ApJ, 539, 413

\bibitem[Kubota \& Makishima(2004)]{2004ApJ...601..428K} Kubota A.,
Makishima K.\ 2004, ApJ, 601, 428

\bibitem[Kubota \& Done(2004)]{2004MNRAS.353..980K} Kubota A., Done
C.\ 2004, MNRAS, 353, 980

\bibitem[Kubota et al.(2001)]{2001ApJ...560L.147K} Kubota A., Makishima
K., Ebisawa K.\ 2001, ApJ, 560, L147

\bibitem[Lee et al.(2002)]{2002ApJ...567.1102L} Lee J.~C., Reynolds
C.~S., Remillard R., Schulz N.~S., Blackman E.~G., Fabian A.~C.\ 2002,
ApJ, 567, 1102

\bibitem[Li et al.(2005)]{2005ApJS..157..335L} Li L.-X., Zimmerman E.~R.,
Narayan R., McClintock J.~E.\ 2005, ApJs, 157, 335

\bibitem[Lubi{\'n}ski \& Zdziarski(2001)]{2001MNRAS.323L..37L}
Lubi{\'n}ski P., Zdziarski A.~A.\ 2001, MNRAS, 323, L37

\bibitem[]{}Magdziarz P., Zdziarski A., 1995, MNRAS, 273, 837

\bibitem[]{} McClintock J.E., Remillard R.A., 2006, in Lewin W.H.G. and van der Klis M. eds,
Compact Stellar X-ray Sources, Cambridge Univ. Press, in press

\bibitem[\protect\citeauthoryear{McConnell et
al.}{2000}]{2000ApJ...543..928M} McConnell M.~L., et al., 2000, ApJ,
543, 928

\bibitem[Miller et al.(2002)]{2002ApJ...570L..69M} Miller J.~M., et al.\
2002, ApJ, 570, L69

\bibitem[Miller et al.(2004)]{2004ApJ...601..450M} Miller J.~M., et al.\
2004a, ApJ, 601, 450

\bibitem[]{} Miller J.M., Fabian A.C., Nowak M.A., Lewin W.H.G.,
  2004b, to appear in proc. 10th Annual Marcel Grossmann Meeting on General Relativity,
(astro-ph/0402101)

\bibitem[\protect\citeauthoryear{Miniutti, Fabian, \&
Miller}{2004}]{2004MNRAS.351..466M} Miniutti G., Fabian A.~C., Miller
J.~M., 2004, MNRAS, 351, 466

\bibitem[]{} Narayan R., Yi., I 1995, ApJ, 452, 710

\bibitem{} Nayakshin S., Kazanas D., Kallman T. R., 2000, ApJ, 537, 833

\bibitem[Orosz et al.(2004)]{2004ApJ...616..376O} Orosz J.~A., McClintock
J.~E., Remillard R.~A., Corbel S.\ 2004, ApJ, 616, 376

\bibitem[Pounds et al.(2003)]{2003MNRAS.345..705P} Pounds K.~A., Reeves
J.~N., King A.~R., Page K.~L., O'Brien P.~T., Turner M.~J.~L.\ 2003,
MNRAS, 345, 705

\bibitem[\protect\citeauthoryear{Poutanen \&
Coppi}{1998}]{1998tx19.confE.355P} Poutanen J., Coppi P.~S., 1998, in
19th Texas Symposium on Relativistic Astrophysics and Cosmology, eds.
J. Paul, T. Montmerle, and E. Aubourg

\bibitem[Proga \& Kallman(2002)]{2002ApJ...565..455P} Proga D.,
Kallman T.~R.\ 2002, ApJ, 565, 455

\bibitem[]{pn00} Psaltis D., Norman C., 2000, preprint
(astro-ph/0001391)

\bibitem[\protect\citeauthoryear{Reynolds \&
Begelman}{1997}]{1997ApJ...488..109R} Reynolds C.~S., Begelman M.~C.,
1997, ApJ, 488, 109

\bibitem[Ross \& Fabian(1993)]{1993MNRAS.261...74R} Ross R.~R., Fabian
A.~C.\ 1993, MNRAS, 261, 74

\bibitem{} Ross R. R., Fabian A. C., Young A. J., 1999, 306, 461

\bibitem[Rossi et al.(2004)]{2004NuPhS.132..416R} Rossi, S., Homan, J.,
Miller, J.~M., \& Belloni, T.\ 2004, Nuclear Physics B Proceedings
Supplements, 132, 416

\bibitem[]{rc00} R{\'o}{\.z}a{\'n}ska A., Czerny B., 2000, MNRAS, 316, 473

\bibitem[]{} Shafee R., McClintock J.E., Narayan R., Davis S.W., Li
  L.--X., Remillard R.A., 2005, ApJL, submitted (astro-ph/0508302)

\bibitem[]{} Shakura N. I., Sunyaev R. A., 1973, A\&A, 24, 337

\bibitem[Shapiro et al.(1976)]{1976ApJ...204..187S} Shapiro S.~L.,
Lightman A.~P., Eardley D.~M.\ 1976, ApJ, 204, 187

\bibitem[\protect\citeauthoryear{Stella \&
Vietri}{1998}]{1998ApJ...492L..59S} Stella L., Vietri M., 1998, ApJ,
492, L59

\bibitem[Titarchuk et al.(1999)]{1999ApJ...525L.129T} Titarchuk L.,
Osherovich V., Kuznetsov S.\ 1999, ApJ, 525, L129

\bibitem[Tomsick et al.(2003)]{2003ApJ...592.1100T} Tomsick J.~A.,
Kalemci E., Corbel S., Kaaret P.\ 2003, ApJ, 592, 1100

\bibitem[Ueda et al.(1998)]{1998ApJ...492..782U} Ueda Y., Inoue H.,
Tanaka Y., Ebisawa K., Nagase F., Kotani T., Gehrels N.\ 1998, ApJ,
492, 782

\bibitem[Ueda et al.(2004)]{2004ApJ...609..325U} Ueda Y., Murakami H.,
Yamaoka K., Dotani T., Ebisawa K.\ 2004, ApJ, 609, 325

\bibitem[Volonteri et al.(2005)]{2005ApJ...620...69V} Volonteri M., Madau
P., Quataert E., Rees M.~J.\ 2005, ApJ, 620, 69

\bibitem[]{wk99} Wijnands R., van der Klis M., 1999, ApJ, 514, 939

\bibitem[\protect\citeauthoryear{Wilms et al.}{2001}]{2001MNRAS.328L..27W}
Wilms J., Reynolds C.~S., Begelman M.~C., Reeves J., Molendi S.,
Staubert R., Kendziorra E., 2001, MNRAS, 328, L27

\bibitem[Young et al.(2005)]{2005ApJ...631..733Y} Young A.~J., Lee J.~C.,
Fabian A.~C., Reynolds C.~S., Gibson R.~R., Canizares C.~R.\ 2005, ApJ,
631, 733

\bibitem[]{zg04} Zdziarski A.A., Gierli{\'n}ski M., 2004, Prog.Theor.Phys.Suppl., 155, 99

\bibitem[]{} {\.Z}ycki P.T., Done C., Smith D.A., 1998, ApJ, 496, L25

\end{thebibliography}
\end{document}